\documentclass[journal]{IEEEtran}
\ifCLASSINFOpdf
\else
\fi

\usepackage{graphicx}
\usepackage{color}
\usepackage{placeins}
\usepackage{float}
\usepackage{epstopdf}
\usepackage[caption = false]{subfig}
\usepackage{tabularx,colortbl}

\begin{document}
\title{A Multiband Patch Antenna with Co-Located Phase Centers for High-Accuracy Inter-Node Ranging in Distributed Antenna Arrays}
\author{John~Doroshewitz,~\IEEEmembership{Student~Member,~IEEE,}
		Christopher~Oakley,~\IEEEmembership{Student~Member,~IEEE,}
		Vincens~Gjokaj,~\IEEEmembership{Student~Member,~IEEE,}
        and~Jeffrey~A.~Nanzer,~\IEEEmembership{Senior~Member,~IEEE}

\thanks{Manuscript received March 2019.}
\thanks{The authors are with with the Electrical and Computer Engineering Department, Michigan State University, East Lansing, MI 48824 (e-mail: nanzer@msu.edu).}}

\markboth{IEEE}%
{Shell \MakeLowercase{\textit{et al.}}: Bare Demo of IEEEtran.cls for Journals}

\maketitle

\begin{abstract}
\boldmath
A unique three-band patch antenna for high-accuracy ranging between nodes in open-loop coherent distributed antenna arrays is presented. Open-loop coherent distributed operations require accurate relative position knowledge between nodes to enable distributed beamforming operations. For the highest accuracy, the inter-node ranging will typically occur at frequencies higher than the frequency of the signal transmitted by the distributed array (the coherent action signal) to enable wider bandwidth signals to be used, however it is important that the inter-node ranging measure the distance between the antennas transmitting the coherent action signal. In this work, a three-band antenna is presented that is designed to support distributed beamforming at 1.88 GHz and high-accuracy ranging using a sparse, two-tone waveform operating near 9.5 and 10.5 GHz. The two-tone waveform is supported by a slotted patch antenna surrounded by a larger patch antenna supporting the 1.88 GHz array signal. The antennas are concentrically designed to ensure that the phase centers of the ranging antenna and the coherent action antenna are closely aligned. Simulated and measured performance shows phase center displacement of approximately $\lambda$/10  relative to the coherent action signal, while maintaining $\mathrm{S_{11}}$ below -10 dB at each band.

\end{abstract}

\IEEEpeerreviewmaketitle

\section{Introduction}


\IEEEPARstart{T}{he disaggregation} of large, single-aperture antennas and phased arrays into systems of coordinated individual wireless systems operating as distributed arrays enables a broad range of significant advances in wireless technologies. Not only does disaggregation make the wireless system more reliable and robust to interference and failures, such distributed arrays can achieve equivalent or greater performance in both gain and spatial resolution than single-platform wireless systems \cite{4102876}. Furthermore, the spatial adaptivity afforded by distributed arrays supports adaptive methods previously unavaliable to spatially-limited single-platform systems. Beamforming from a distributed array can be achieved using feedback from the destination location when transmitting to or receiving from a cooperative wireless system (e.g. \cite{4202181}), however the greatest benefits from distributed beamforming are achieved when the array is implemented in an open-loop format, where no feedback from the destination is available \cite{7803582}.

Open-loop distributed beamforming requires the signals emitted by the individual nodes to be coordinated at the wavelength level of the transmitted waveform. To ensure that phase-coherent distributed beamforming takes place requires the frequencies of the signals to be locked and the relative phases to be appropriately set such that the waveform adds coherently in the desired beamforming direction. The most challenging requirements for phase alignment is determining the relative distances between the transmitting antennas on separate, potentially moving platforms \cite{7118937}. In particular, it is important that the relative distances between the antennas performing the distributed action be known to a distance of less than a wavelength to ensure that phase errors do not significantly degrade the coherent gain of the distributed beamforming operation. 

In this work, a unique multiband antenna designed to enable high-accuracy ranging between the phase centers of the beamforming antennas on the nodes in a distributed beamforming array is presented. The ranging antenna is based around a novel spectrally-sparse, high-accuracy ranging waveform, with multiple resonances tuned to the specific frequencies in the ranging waveform. This antenna is surrounded concentrically by a lower frequency antenna designed for the distributed beamforming operation. By aligning the two antennas concentrically, the phase centers are physically co-located. Simulated and measured performance shows phase center displacement of less than one-third of a wavelength of the coherent action signal, while maintaining $\mathrm{S_{11}}$ of below -10 dB and positive gain at each band.

\section{Antenna Design for Spectrally-Sparse Ranging Waveforms}

The accuracy of estimating the range is given by the variance in estimating the time of arrival $\tau$ of a signal reflected back from the second node,
\begin{equation}
\mathrm{var}(\hat{\tau}-\tau)\geq \frac{N_0}{2|\alpha|^2\zeta_f^2}
\end{equation}
where $N_0$ is the noise amplitude, $\alpha$ is the amplitude of the signal, and $\zeta_f$ is the mean-squared bandwidth. This variance is inversely proportional to the mean squared bandwidth as well as the signal-to-noise ratio (SNR). In order to have the most accurate estimation possible, this variance should be minimized. Increasing the mean squared bandwidth of the signal will reduce this variance. The mean squared bandwidth is calculated by
\begin{equation}
\zeta_f^2=\int(2\pi f)^2|G(f)|^2df
\end{equation}
where $G(f)$ is the spectral content of the signal. Concentrating the energy into the sidebands of the wideband signal increases this mean squared bandwidth while preserving the same usable bandwidth of the system. Use of this two-tone signal not only improves the accuracy of the range estimation, it is often easier to generate these types of signals in hardware, compared to one wideband signal across the same bandwidth. Previous work \cite{7118937} has shown that adequate ranging accuracy of a millimeter or better can be achieved at the X-band using a 500 MHz tone separation. This work, however, was performed at 16 dB SNR. In future distributed radar applications a much lower SNR environment is anticipated. In order to account for this lower SNR, a tone separation of approximately 1 GHz is desired to keep millimeter accuracy in the range estimation.

Based on this analysis, the best ranging waveform consists of two separate narrow signals separated by a wide bandwidth. The antenna supporting is thus best designed to include two narrow resonances matching these tones, while filtering out any signals in-between to minimize the noise contributions. Combined with this frequency, a third band is also present where the coherent distributed beamforming takes place. The antenna should thus support three separate bands simultaneously. 
Co-locating these antennas in one multiband antenna will give the most accurate ranging between the antennas implementing the distributed beamforming. Knowing that these phase centers at each frequency originate from approximately the same location allows the range estimation to be used for the coherent action frequency with reasonable assurance that it is accurate. In addition to the co-located phase centers, a multiband antenna improves the size and weight of the hardware required per node in the distributed array. Since this array is envisioned to be on mobile platforms, the physical improvements will help make an overall more efficient system.

The microstrip patch antenna is of particular interest in this application due to its small size, low-cost, ease of fabrication, and low profile. Traditionally, patch antennas have been designed to operate efficiently at one frequency, however significant effort has been focused on wideband and multiband operation. For many years, microstrip antennas have been designed for multiband operation while still preserving minimal antenna height \cite{1187424}, \cite{1321323} and evolutionary algorithms have been employed to optimize multiband and wideband patch antennas \cite{1528712}, \cite{7999969}. Recent work in multiband microstrip antennas has focused on operation for WLAN, WiMAX, LTE and other common communication standards \cite{7805264}, \cite{7479524}.

\section{Multiband Antenna Design}
The antenna design was performed in three steps. First, a slotted patch antenna was designed to handle the two upper frequencies for the two-tone ranging waveform. Next, a partial patch was added surrounding the slotted patch antenna that is used for the low frequency coherent action. Lastly, microstrip stub filters were added to the design to isolate the currents into the proper areas of the antenna. 


The primary radiator for the two-tone frequencies uses a slotted patch antenna. This design has been shown previously to be able to create two resonant frequencies simultaneously \cite{646798}. Not only does this have the capability to provide two resonances at approximately a 1 GHz separation, there is very little resonance between these two tones. This provides natural noise rejection capabilities without the introduction of extra filters. The noise rejection will help the overall SNR of the system, which will increase the accuracy of the range estimation.
%
The slotted patch design was integrated into the overall design of the three-band antenna. The lower frequency resonance was created by adding copper surrounding the previously designed slotted patch antenna. A large rectangle with the center removed was produced. This created a partial patch antenna, with similar radiation characteristics to a traditional patch antenna. Although this produces some parasitic effects on the inner slotted patch antenna, the changes in radiation characteristics were small and it preserved its sensitive design. Minor changes to the dimensions of the slotted patch were made to account for these changes.

\begin{figure}[t]
\begin{center}
\noindent
	\includegraphics[width=3.4in]{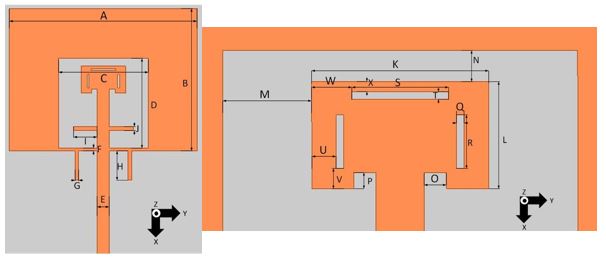}
	\caption{Full patch geometry, dimensions can be seen in Table \ref{dimensions}.}\label{full_patch}
\end{center}
\end{figure}

\begin{figure}[t]
\begin{center}
\noindent
	\includegraphics[width=3.4in]{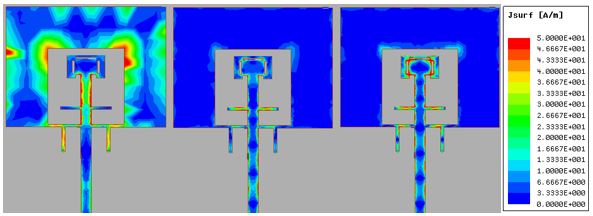}
	\caption{Simulated current distributions demonstrating the successful operation of the filtering stubs. Densities are at the three main simulated frequencies of 1.89 GHz (left), 9.45 GHz (middle), and 10.70 GHz (right).}\label{currents}
\end{center}
\end{figure}

In order to improve the operation of the antenna, stub filtering was used to isolate the currents into the correct areas of the multiband antenna. On the small 100 $\Omega$ transmission lines feeding into the partial patch antenna shorted stub filters were added. These stubs act as a low pass filter to keep the high frequency currents out of this area. On the 50 $\Omega$ transmission line feeding into the slotted patch antenna, open stubs were added. These stubs act as a high pass filter to isolate the high frequency currents into the slotted patch antenna area. These three features were combined to create the overall antenna design, which can be seen in Fig. \ref{full_patch}. The overall dimensions can be seen in Table \ref{dimensions}. The simulated current distributions at each of the three bands can be seen in Fig. \ref{currents}. 

\begin{table}[t]
\begin{center}
\caption{Multiband Patch Antenna Dimensions (mm)}\label{dimensions}
\begin{tabular}{ |c c | c c | c c |  }
 \hline
 A & 51.5 & I & 6.5 & Q & 0.7 \\
 B & 38.83 & J & 1 & R & 4.1 \\
 C & 24.5 & K & 12.2 & S & 6.5 \\
 D & 24.5 & L & 7.4 & T & 0.5 \\
 E & 3.34 & M & 6.15 & U & 1.7 \\
 F & 0.83 & N & 2.155 & V & 1.6 \\
 G & 1 & O & 1.68 & W & 2.85 \\
 H & 8 & P & 1.1 & X & 0.7 \\
 \hline
\end{tabular}
\end{center}
\end{table}

\section{Performance Analysis}
\begin{figure}[t]
\begin{center}
\noindent
	\includegraphics[width=1.75in]{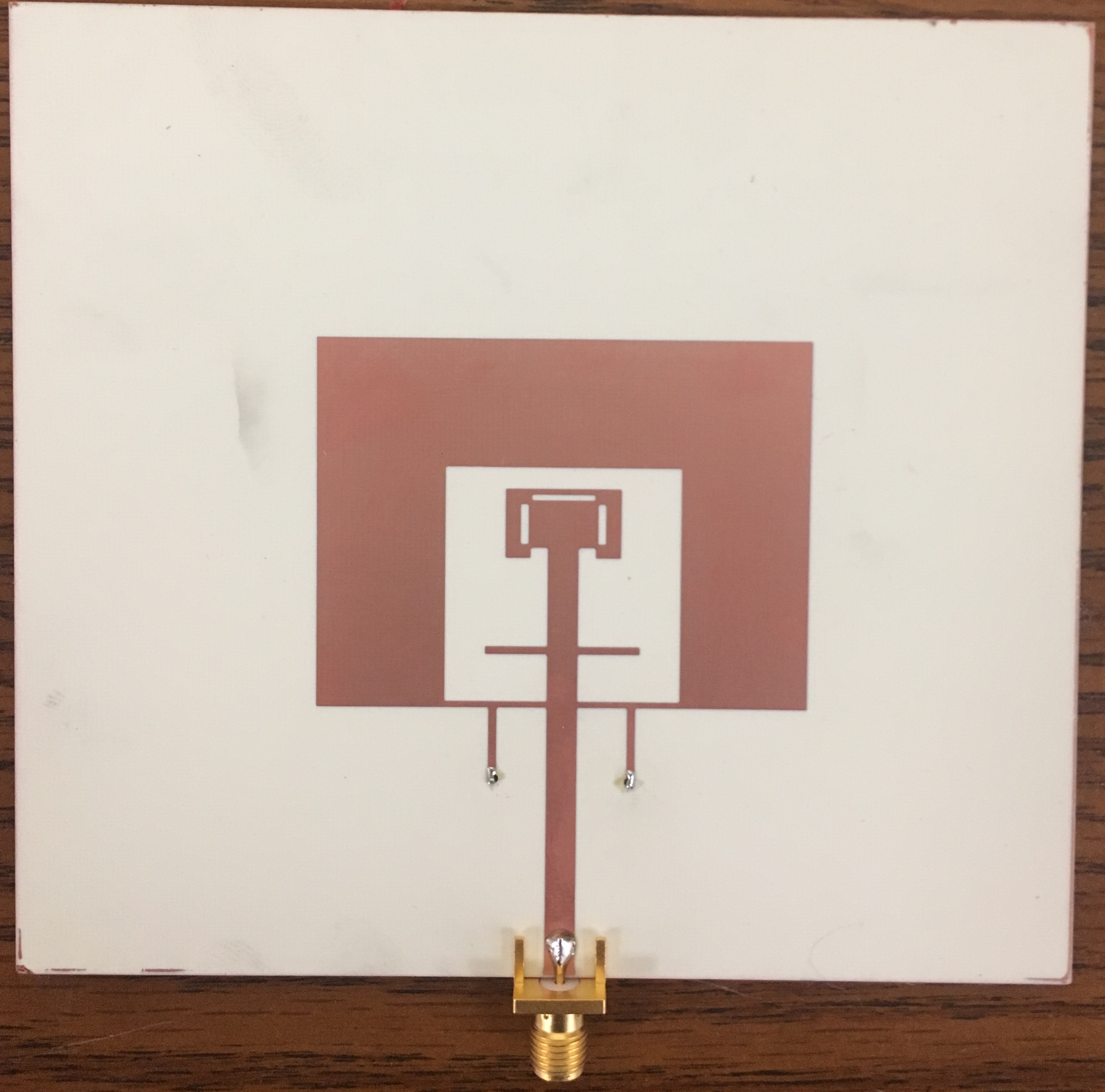}
	\caption{Fabricated multiband patch antenna on Rogers 4350B.}\label{fab}
\end{center}
\end{figure}

The antenna was fabricated on Rogers 4350B dielectric board. This board has a dielectric constant of $\epsilon_r\approx3.66$ and a height of 1.524 mm. Holes were drilled at the end of the shorted stubs and small wires were soldered to the ground plane on the rear of the antenna. This technique does provide some unwanted parasitics to the antenna but it is generally reliable at the frequencies for which it was designed. An end-launch SMA connector was added and the entire ground plane was preserved on the back side of the antenna. The total size of the substrate was 110 mm by 100 mm. The fabricated antenna can be seen in Fig. \ref{fab}.

\begin{figure}[t]
\begin{center}
\subfloat{\includegraphics[width=3.25in]{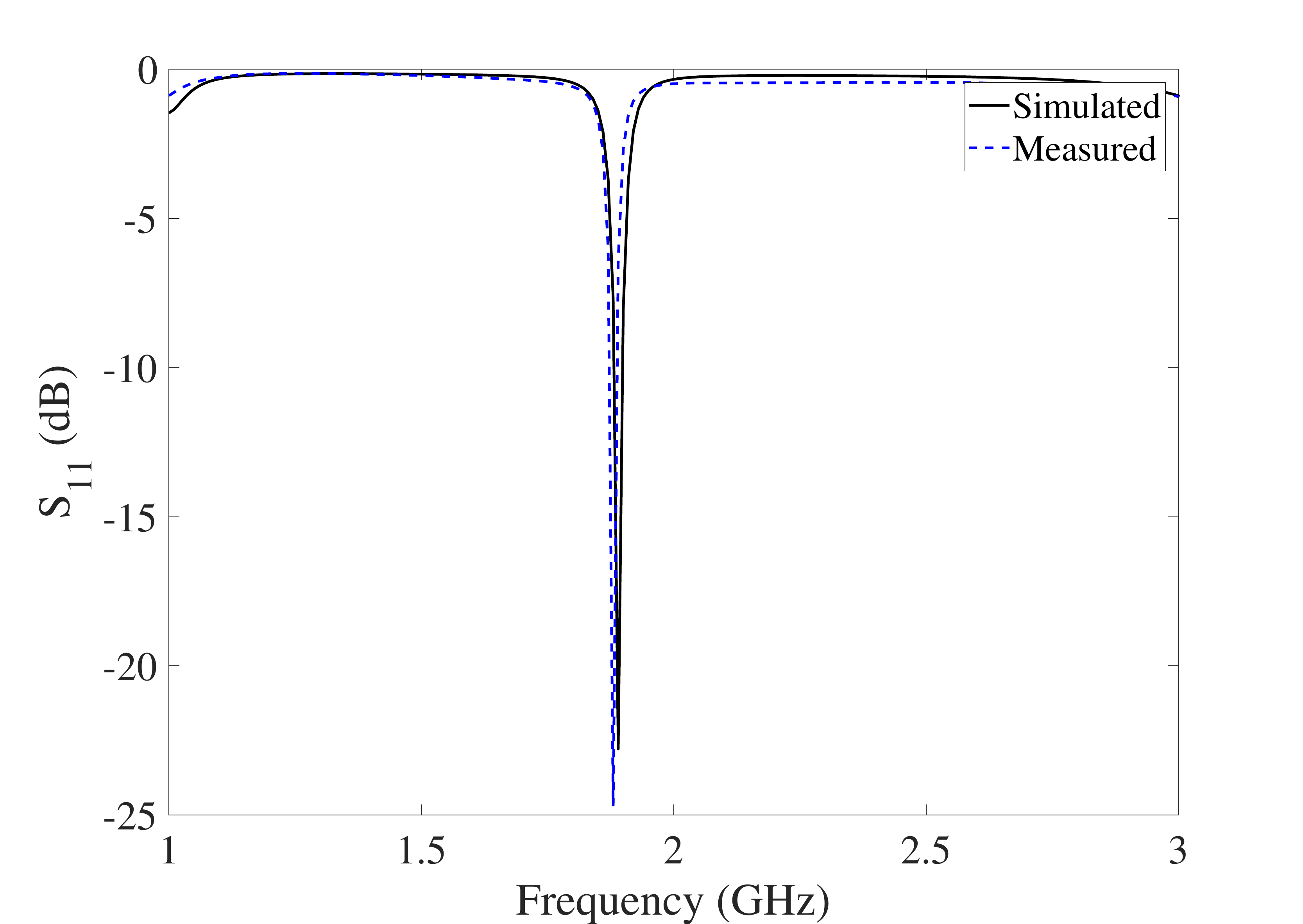}}\\
\subfloat{\includegraphics[width=3.25in]{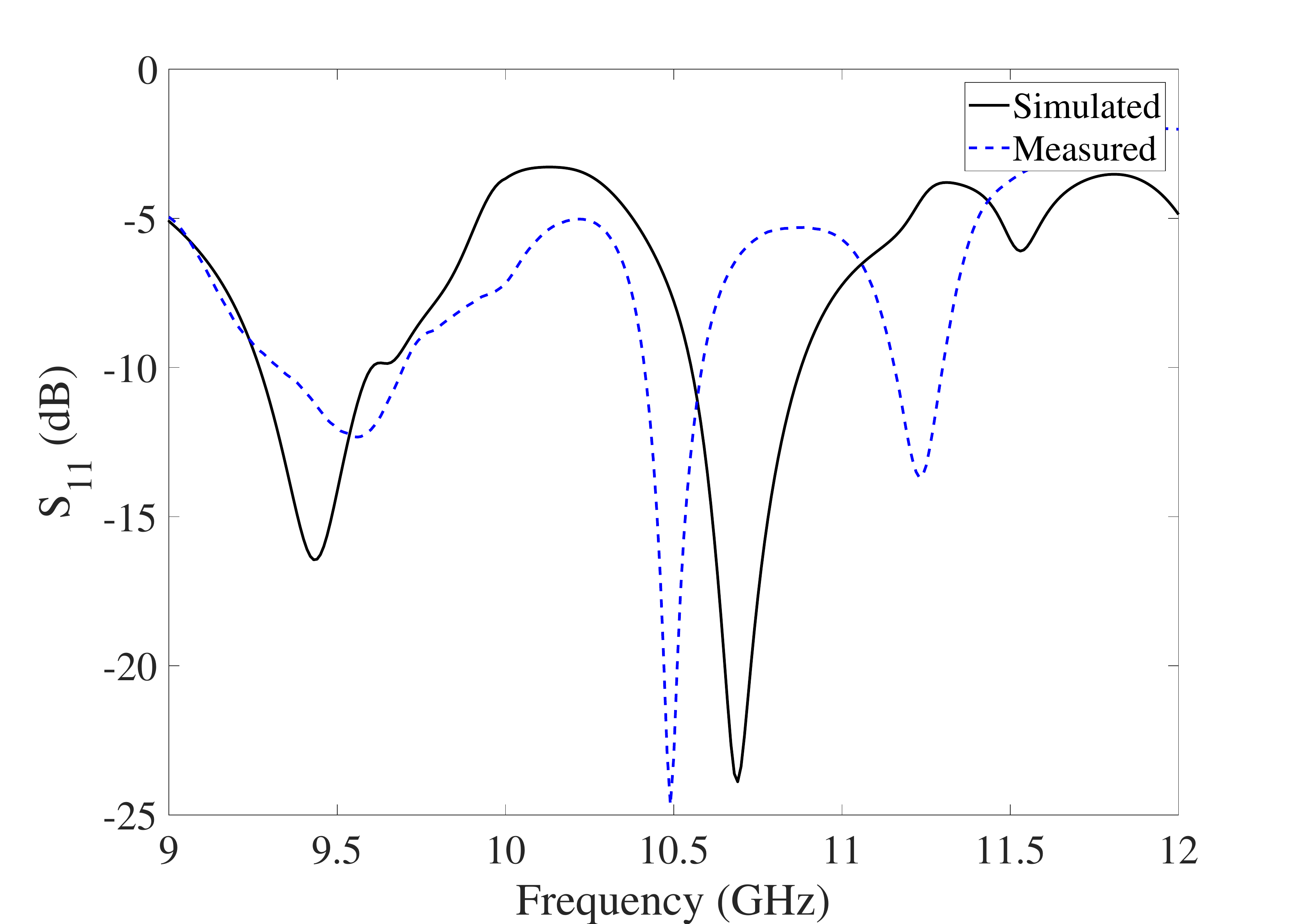}}
\caption{Simulated and measured $S_{11}$ across the frequency bands of interest.}
\label{full_patch_s11}
\end{center}
\end{figure}

The $\mathrm{S_{11}}$ of the fabricated antenna was measured using an Agilent E5071C network analyzer. The measured $\mathrm{S_{11}}$ can be seen in Fig. \ref{full_patch_s11}. The measured low frequency resonance matched well with simulation. At the two upper frequencies, the measured $\mathrm{S_{11}}$ trended well with simulation, but there was a larger frequency shift at both tones than what was seen at the low frequency. Additionally, one tone shifted up in frequency from simulation and the other tone shifted down in frequency. This was not only due to manufacturing variance in the dielectric constant of the substrate, which is normal and expected in most microstrip antenna design, but also to the sensitive nature of the slotted patch antenna. During the design of both the slotted patch and the full patch antenna, it was observed that the slots are extremely sensitive to small changes in their dimensions. The slight over-etching of each dimension of the slotted patch collectively affected the current paths to show a shift in resonance of both of the upper tones. In addition, the use of soldered shorting wires is different than simulation, which could contribute to the shifts of resonance at the upper frequencies.

\begin{figure}[t]
\begin{center}
\noindent
\includegraphics[width=2.5in]{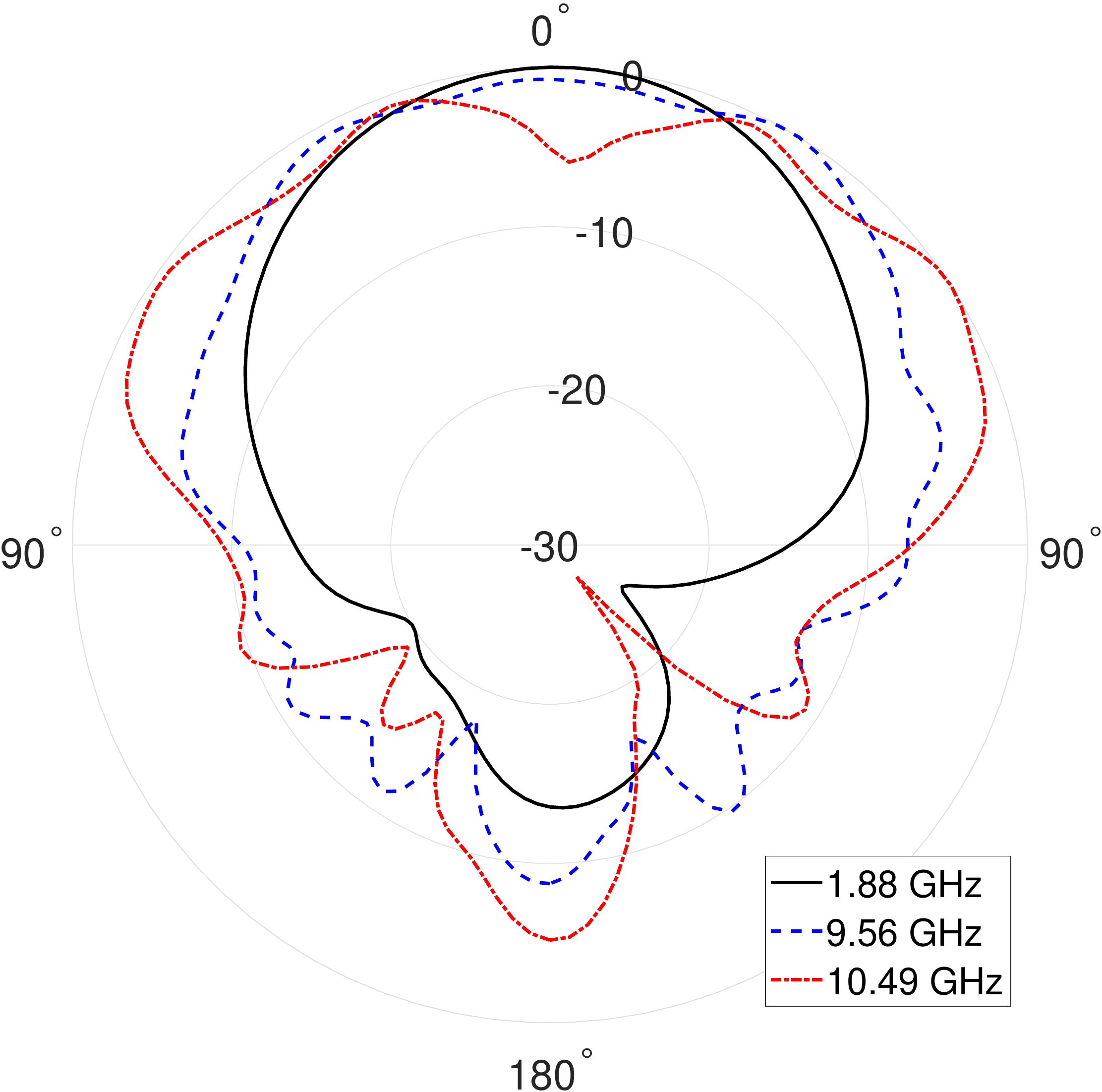}
\caption{Normalized radiations patterns ($\phi=0^\circ$ cut) at each frequency band.}
\label{patterns}
\end{center}
\end{figure}

The low frequency resonance was measured at 1.88 GHz with a fractional bandwidth under -10 dB of 1.06\%. The measured $\mathrm{S_{11}}$ at this frequency was -24.70 dB. The two-tone frequencies were measured at 9.56 GHz and 10.49 GHz, with fractional bandwidths under -10 dB of 3.97\% and 1.71\%, respectively. The measured $\mathrm{S_{11}}$ at the middle frequency was -12.26 dB, and -24.67 dB at the highest frequency. The normalized gain patterns at each frequency band can be seen in Fig. \ref{patterns}. These patterns were measured using a Satimo StarLab near-field antenna measurement system. The maximum and average gains can be seen in Table \ref{gain_table}. Since each frequency in practice is transmitting simultaneously, only the gain in the region $60^\circ$ above the antenna is presented.


\begin{table}[t]
\begin{center}
\caption{Antenna gain in $60^\circ$ main beam at each frequency band}\label{gain_table}
\begin{tabular}{ | c | c | c | }
 \hline
 \hspace{.1in} & Max & Mean\\
 \hline \hline
 1.88 GHz & 4.8003 dB & 3.9011 dB\\
 9.56 GHz & 1.9413 dB & 1.0790 dB\\
 10.49 GHz & 3.5768 dB & 1.8261 dB\\
 \hline
\end{tabular}
\end{center}
\end{table}

\begin{figure}[t]
\begin{center}
{\includegraphics[width=3.5in]{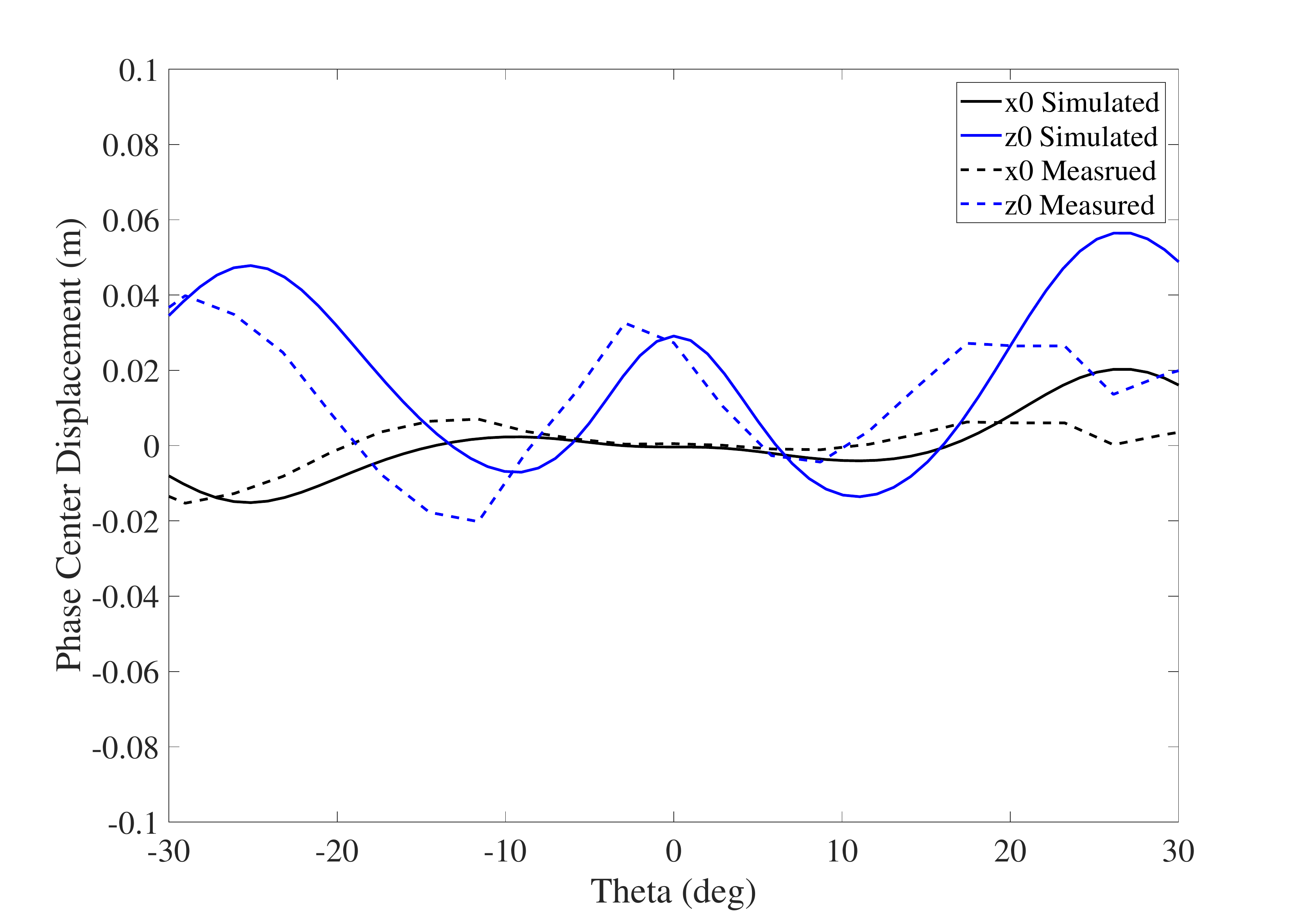}}

(a)

{\includegraphics[width=3.5in]{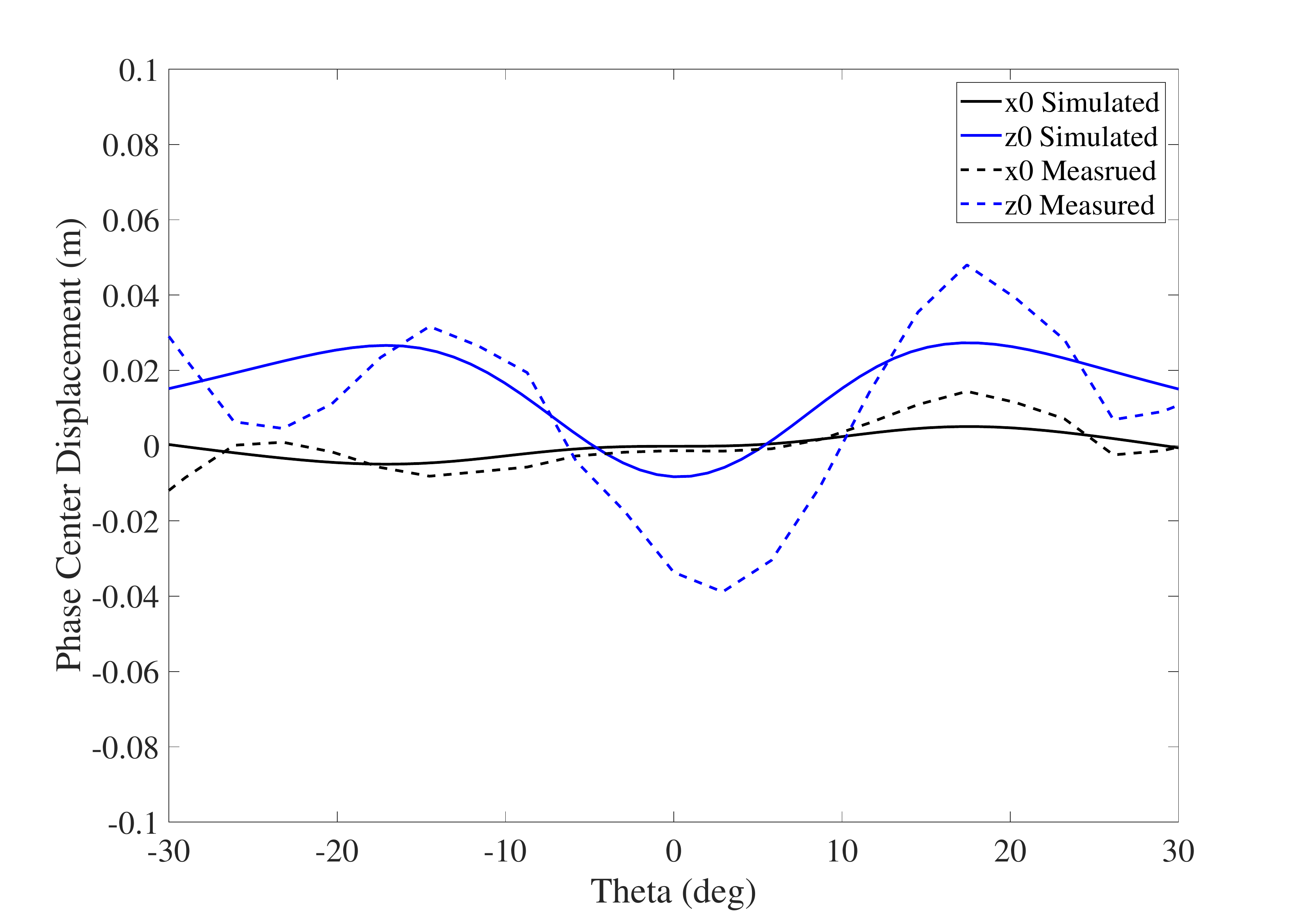}}

(b)

\caption{Difference of the displacement of the phase center between (a) the high and low frequencies and (b) the middle and low frequencies in the main beam of the $\phi=0^\circ$ cut.}
\label{phase_center_0}
\end{center}
\end{figure}

Because this antenna is intended to be used for high accuracy distance estimation, a calculation of the phase center at each frequency is necessary. The difference in displacement of the phase centers between each frequency at the $\phi=0^\circ$ cut can be seen in Fig. \ref{phase_center_0}.  These displacements were calculated using the equations described in \cite{prata2002misaligned}. The mean and standard deviation of the phase center (Table \ref{phase_center}) shows that the total displacement between each band is approximately $\lambda$/10 at the 1.88 GHz coherent action frequency.


\begin{table}[t!]
\begin{center}
\caption{Phase Center Displacement Difference Mean and Standard Deviation (in meters)}\label{phase_center}
\begin{tabular}{ | c | c c c c |  }
 \hline
  & x0 Mean & x0 SD & z0 Mean & z0 SD\\
 \hline \hline
 Low-High & 0.0002 & 0.0062 & 0.0141 & 0.0168\\
 Low-Mid & -0.0006 & 0.0073 & 0.0107 & 0.0241\\
 \hline
\end{tabular}
\end{center}
\end{table}

\section{Conclusion}
A three band antenna for use in distributed arrays was presented. This antenna can handle a low frequency resonance for coherent action in the distributed array while simultaneously transmitting a two-tone signal intended for ranging between nodes in the array. The antenna consists of a slotted patch antenna for the two-tone signal that is surrounded by a partial patch antenna for the lower frequency. Proper measured $\mathrm{S_{11}}$ was observed and reasonable gain and patterns were presented at each of the three frequency bands.


\ifCLASSOPTIONcaptionsoff
  \newpage
\fi

%
%
%
%

\bibliographystyle{IEEEtran}
\bibliography{IEEEabrv,multiband_patch_rev2}{}

\begin{thebibliography}{10}
\providecommand{\url}[1]{#1}
\csname url@samestyle\endcsname
\providecommand{\newblock}{\relax}
\providecommand{\bibinfo}[2]{#2}
\providecommand{\BIBentrySTDinterwordspacing}{\spaceskip=0pt\relax}
\providecommand{\BIBentryALTinterwordstretchfactor}{4}
\providecommand{\BIBentryALTinterwordspacing}{\spaceskip=\fontdimen2\font plus
\BIBentryALTinterwordstretchfactor\fontdimen3\font minus
  \fontdimen4\font\relax}
\providecommand{\BIBforeignlanguage}[2]{{%
\expandafter\ifx\csname l@#1\endcsname\relax
\typeout{** WARNING: IEEEtran.bst: No hyphenation pattern has been}%
\typeout{** loaded for the language `#1'. Using the pattern for}%
\typeout{** the default language instead.}%
\else
\language=\csname l@#1\endcsname
\fi
#2}}
\providecommand{\BIBdecl}{\relax}
\BIBdecl

\bibitem{4102876}
R.~C. Heimiller, J.~E. Belyea, and P.~G. Tomlinson, ``Distributed array
  radar,'' \emph{IEEE Transactions on Aerospace and Electronic Systems}, vol.
  AES-19, no.~6, pp. 831--839, Nov 1983.

\bibitem{4202181}
R.~Mudumbai, G.~Barriac, and U.~Madhow, ``On the feasibility of distributed
  beamforming in wireless networks,'' \emph{IEEE Transactions on Wireless
  Communications}, vol.~6, no.~5, pp. 1754--1763, May 2007.

\bibitem{7803582}
J.~A. Nanzer, R.~L. Schmid, T.~M. Comberiate, and J.~E. Hodkin, ``Open-loop
  coherent distributed arrays,'' \emph{IEEE Transactions on Microwave Theory
  and Techniques}, vol.~65, no.~5, pp. 1662--1672, May 2017.

\bibitem{7118937}
J.~E. Hodkin, K.~S. Zilevu, M.~D. Sharp, T.~M. Comberiate, S.~M. Hendrickson,
  M.~J. Fitch, and J.~A. Nanzer, ``Microwave and millimeter-wave ranging for
  coherent distributed rf systems,'' in \emph{2015 IEEE Aerospace Conference},
  March 2015, pp. 1--7.

\bibitem{1187424}
K.-L. Wong, G.-Y. Lee, and T.-W. Chiou, ``A low-profile planar monopole antenna
  for multiband operation of mobile handsets,'' \emph{IEEE Transactions on
  Antennas and Propagation}, vol.~51, no.~1, pp. 121--125, Jan 2003.

\bibitem{1321323}
Y.-X. Guo, M.~Y.~W. Chia, and Z.~N. Chen, ``Miniature built-in multiband
  antennas for mobile handsets,'' \emph{IEEE Transactions on Antennas and
  Propagation}, vol.~52, no.~8, pp. 1936--1944, Aug 2004.

\bibitem{1528712}
N.~Jin and Y.~Rahmat-Samii, ``Parallel particle swarm optimization and finite-
  difference time-domain (pso/fdtd) algorithm for multiband and wide-band patch
  antenna designs,'' \emph{IEEE Transactions on Antennas and Propagation},
  vol.~53, no.~11, pp. 3459--3468, Nov 2005.

\bibitem{7999969}
V.~Gjokaj, J.~Doroshewitz, J.~Nanzer, and P.~Chahal, ``A design study of 5g
  antennas optimized using genetic algorithms,'' in \emph{2017 IEEE 67th
  Electronic Components and Technology Conference (ECTC)}, May 2017, pp.
  2086--2091.

\bibitem{7805264}
J.~Dong, X.~Yu, and L.~Deng, ``A decoupled multiband dual-antenna system for
  wwan/lte smartphone applications,'' \emph{IEEE Antennas and Wireless
  Propagation Letters}, vol.~16, pp. 1528--1532, 2017.

\bibitem{7479524}
M.~van Rooyen, J.~W. Odendaal, and J.~Joubert, ``High-gain directional antenna
  for wlan and wimax applications,'' \emph{IEEE Antennas and Wireless
  Propagation Letters}, vol.~16, pp. 286--289, 2017.

\bibitem{646798}
S.~Maci and G.~B. Gentili, ``Dual-frequency patch antennas,'' \emph{IEEE
  Antennas and Propagation Magazine}, vol.~39, no.~6, pp. 13--20, Dec 1997.

\bibitem{prata2002misaligned}
A.~Prata, ``Misaligned antenna phase-center determination using measured phase
  patterns,'' \emph{IPN Progress Report}, pp. 42--150, 2002.

\end{thebibliography}

\end{document}